\ifx\submissionver \undefined
\documentclass{article}

\else
\documentclass{llncs}
\fi

\usepackage{array, latexsym, graphicx,times,amsfonts,picture,color,multirow,epsf,epsfig,mathtools}
\usepackage{amsmath}
\usepackage{verbatim}
\usepackage{blkarray}
\usepackage{setspace}
\usepackage{breqn}
\usepackage{listings}
\usepackage{enumitem}
\usepackage[T1]{fontenc}
\usepackage{textcomp}
\usepackage{hyperref}

\usepackage[utf8]{inputenc}

\begin{document}
\title{Implementing Automated Market Makers with Constant Circle}
\ifx\submissionver \undefined
\title{Implementing Automated Market Makers with Constant Ellipse}
\author{Yongge Wang\\ UNC Charlotte}
\else
\author{Yongge Wang}
\institute{
UNC Charlotte, 9201 University City Blvd., NC 28223, USA (\email{yonwang@uncc.edu})
}
\fi


\maketitle

\begin{abstract}
This paper describe the implementation details of constant ellipse based automated market makers (CoinSwap).
A CoinSwap prototype has been implemented at \url{http://coinswapapp.io/} and the source codes are available
at \url{https://github.com/coinswapapp/}. 
\end{abstract}

\section{Introduction}
Decentralized finance (DeFi or open finance) is implemented through smart contracts (DApps)
which are stored on a public distributed ledger (such as a blockchain)
and can be activated to automate execution of financial instruments and digital assets. The immutable 
property of blockchains guarantees that these DApps are also tamper-proof and the content 
could be publicly audited. Recently, Wang \cite{wang2020automated} proposed a constant ellipse/circle
based automated market makers that improves existing schemes such 
as LMSR \cite{hanson2003combinatorial,hanson2007logarithmic}, 
LS-LMSR  \cite{othman2013practical}, and constant product (e.g., Uniswap \cite{uniswapprotocol})  
based automated market makers.

 \section{Constant ellipse/circle automated market makers}
 \label{conscirsec}
Recently, Wang \cite{wang2020automated} proposed automated market makers based on constant 
ellipse/circle cost functions. That is, the automated market maker's cost function is defined by 
\begin{equation}
\label{constantcircleeq}
C({\bf q})={\sum_{i=1}^n (q_i-a)^2+b\sum_{i\not=j}q_iq_j}
\end{equation}
where $a, b$ are constants.  In constant ellipse/circle automated market makers, the price function for each token is 
$$P_i({\bf q})=\frac{\partial C({\bf q})}{\partial q_i}={2(q_i-a)+b\sum_{j\not=i}q_j}.$$
For automated market makers, we only use the first quadrant of the coordinate plane.
By adjusting the parameters $a,b$ in the equation (\ref{constantcircleeq}),
one may keep the cost function to be concave (that is, using the upper-left part of the ellipse/circle)
or to be convex (that is, using the lower-left part of the ellipse/circle). 

For automated market makers based on the cost function $C({\bf q})=(x-10^{9})^2 + (y-10^{9})^2$, if the initial
deposit to the market is $10^{6}$ coins (if the decimal value for the token is 18, then this is equivalent to $10^{24}$ 
token unit)  for each token, then the market cost is 
$C({\bf q}_0)=2\cdot 10^{12}\cdot 999^2.$ 
At market state ${\bf q}_0$, a trader may use $1001002$ token $B$ coins
to purchase  1000000 token $A$ coins 
with a resulting market state ${\bf q}_1=(0, 2001002)$ and a resulting market cost $C({\bf q}_1)=C({\bf q}_0)$.
This tangent line slope stays in the interval $[-1.002005014, -0.9979989980]$. Note that the boundary 
price is reached when the balance of one token becomes zero. Assume that the cost function is $r \cdot 10^{14}=(x-10^{9})^2 + (y-10^{9})^2$
and $x=0$. Then the price of the first token reaches the highest value of 
$$\frac{\partial y}{\partial x}=\frac{10^9}{\sqrt{r\cdot 10^{14} -10^{18}}}
=\frac{100}{\sqrt{r-1000}}$$
The token price fluctuation could be adjusted by revising the value of the circle radius. The following table shows 
some examples.
\begin{center}
\begin{tabular}{| c| c| c| } \hline
{\bf value }$r$ & minimal price & maximum price \\ \hline
 $10001$&0.01000000000&100.0000000 \\ \hline
 $10010$&0.03162277660&31.62277660 \\ \hline
 $10100$&0.1000000000&10.00000000 \\ \hline
 $10500$&0.2236067977& 4.472135956\\ \hline
 $11000$&0.3162277660&3.162277660\\ \hline
 $15000$&0.7071067814&1.414213562\\ \hline
 $17000$&0.8366600265&1.195228609 \\ \hline
 $19000$&0.9486832980&1.054092553 \\ \hline
 $19900$&0.9949874374&1.005037815 \\ \hline
 $19990$&0.9994998752&1.000500375 \\ \hline
 $19999$&0.9999499985&1.000050004 \\ \hline
\end{tabular}
\end{center}


\section{Implementing constant ellipse based AMM -- Approach I}
\label{implappI}
As an example, we use the circle $(x-c)^2 + (y-c)^2=r^2$
to show how to establish a token pair swapping market in this section. Specifically,
we use $c=10^9$ and $r\cdot 10^{14}=16000\cdot 10^{14}$ (that is, $r=16000$) for illustration purpose in this section.
Each token pair market maintains constants $\lambda_0$ and $\lambda_1$  which are determined 
at the birth of the market. Furthermore, each token market also maintains
a non-negative multiplicative scaling variable $\mu$ which is the minimal value 
so that the following equation holds.
\begin{equation}
\label{1stmu}
(\mu \lambda_0x_0-10^{9})^2 + \left({\mu\lambda_1 y_0}-10^{9}\right)^2\le 16000\cdot 10^{14}
\end{equation}
where $\mu \lambda_0x_0<10^9$ and 
${\mu\lambda_0 y_0}<10^9$. This ensures that we use the lower-left section of the circle
for the automated market.  The tangent line's slope 
at the point $(x,y)$ for the pool circle is 
$$\frac{\partial (\lambda_1y)}{\partial (\lambda_0x)}=-\frac{10^9-\mu\lambda_0 x}{10^9-\mu\lambda_1 y}$$
Thus for the reserves $(x_0, y_0)$, the $(\lambda_0, \lambda_1)$-weighted  relative price of the tokens at this market is
\begin{equation}
\label{priceratioeq}
P_{y/x}(x_0,y_0)=\frac{\lambda_0(10^9-\mu\lambda_0x_0)}{\lambda_1(10^9-\mu\lambda_1y_0)}.
\end{equation}
That is, at market $(x_0, y_0)$, $\Delta_x$ token $A$ coins could be swapped for 
$\Delta_y=\Delta_xP_{y/x}(x_0,y_0)$ token $B$ coins.

Each token market also maintains
a non-negative multiplicative scaling variable $\mu$  and a total liquidity supply amount variable
$\Omega$.
The value of $\Omega$ changes each time when a liquidity 
provider adds liquidity  to or removes liquidity from the market. 
The liquidity value $\Omega$ should be defined in such a way that
the following conditions are satisfied. 
Assume that the current market state is ${\bf q}=(x,y)$ with a market liquidity value 
$\Omega_{\bf q}$. If a customer moves the current market state to a new 
state ${\bf q}'=(e\cdot x,e\cdot y)$, then the new market liquidity value is $\Omega_{{\bf q}'}=e\cdot \Omega_{\bf q}$.
This could be achieved in several ways. For example, Uniswap defines $\Omega(x,y)=\sqrt{xy}$.
In this section, we use the following approach. Let $\pi(x,y)=ax+b y$ for some constants $a,b$.
For a given market state ${\bf q}=(x,y)$ with liquidity value $\Omega$, if one moves the market 
state from ${\bf q}=(x,y)$ to ${\bf q}'=(e\cdot x,e\cdot y)$, then the mew market liquidity for ${\bf q}'$ is defined in such
a way that 
\begin{equation}
\label{lqreq}
\frac{\Omega'}{\Omega}=\frac{\pi(e\cdot x,e\cdot y)}{\pi(x,y)}=e.
\end{equation}
The reader should be aware of the fact that when the market moves on, we normally do not have $\Omega(x,y)=\pi(x,y)$. 
It should also be noted that an alternative function $\pi(x,y)=\sqrt{x^2+y^2}$ could also be used
in the equation (\ref{lqreq}) for certain applications.

Though in several other DEX applications (such as Uniswap), a liquidity provider normally provides 
equivalent values of token A and token B to the market each time, this is generally not true for CoinSwaps.
As an example, assume that the current market state is ${\bf q}=(x_0,y_0)$ with a total supply $\Omega$.
A liquidity provider can add some liquidity $(\delta x_0, \delta y_0)$ for some $\delta>0$ to the market
without changing the current $(\lambda_0, \lambda_1)$-weighted  relative price $\delta=P_{y/x}(x_0,y_0)$ in (\ref{priceratioeq}).
That is, we need to keep 
\begin{equation}
\label{addingeqprice}
-P_{y/x}((1+\delta)x_0,(1+\delta)y_0)=\frac{10^9-\mu'\lambda_0(1+\delta)x_0}{10^9-\mu'\lambda_1(1+\delta)y_0}
= \frac{10^9-\mu\lambda_0x_0}{10^9-\mu\lambda_1y_0}
\end{equation}
where $x=x_0+\Delta_x$, $y=y_0+\Delta_y$,  and $\mu'>\mu$. It is straightforward to show 
that $$\mu'=\frac{\mu}{1+\delta}.$$

As an example, assume that when the market is set up, one token $A$'s value is equivalent to three token $B$'s price.
That is, we set $\lambda_0=3$ and $\lambda_1=1$. Now assume at some time with market 
$(x_0,y_0)$, one token $A$'s value is equivalent 
to two token $B$'s price. That is, 
$$\frac{\partial (\lambda_1y)}{\partial (\lambda_0x)}=-\frac{10^9-3\mu x_0}{10^9-\mu y_0}=-\frac{2}{3}$$
which means $y_0=4.5x_0-\frac{10^9}{2\mu}$. A liquidity provider can select a $\delta>0$ and 
add $(\delta x_0, \delta y_0)$ tokens to the market. Generally do not have $\delta x_0=2\delta y_0$
since $y_0=4.5x_0-\frac{10^9}{2\mu}$ does not guarantee $x_0=2y_0$.

\subsection{Adding/removing liquidity and swapping}
\label{addremoveliquidity}
{\bf Establishing a token pair market}.
When a token pair market is not established yet, a liquidity provider can establish 
the token pair market by depositing $x_0$ coins of token $A$ and $y_0$ coins of token $B$.
The assumption is that the market value of $x_0$ coins  of  token $A$ is equivalent to
the market value of $y_0$ coins of token $B$. The token pair constant variables $\lambda_0, \lambda_1$ are defined
as ${\lambda_0}x_0\sim{\lambda_1}{y_0}$. Note that $\lambda_0, \lambda_1$ are extensively used 
by the swapping algorithm. Thus it is better to have simple/smaller $\lambda_0, \lambda_1$.
This could be calculated by the smart contract using continuous fraction. It is recommended 
that the liquidity providers should provide the values of $\lambda_0, \lambda_1$ when establishing 
a new market (or by vote from the community). Let $\mu$ be the non-negative number 
such that the equation (\ref{1stmu}) holds. Let the liquidity of the token pair market be defined as 
$\Omega_0=10^{18}\cdot \frac{\lambda_0x_0+{\lambda_1y_0}}{2}$. 
As a return, the liquidity provider receives $\Omega_0$ liquidity coins for this token pair market. 
In our implementation, it is required that each liquidity provider should put at least 
one coin for each token in the market. That is, for an ERC token with $10^{18}$ decimals,
the liquidity provider should put at least $10^{18}$ units of token $A$ and $10^{18}$ units of token $B$.
This requirement insures that $\mu\le 10^9.$

\vskip 5pt
\noindent
{\bf Data representations:} We have the following conventions:
\begin{itemize}
\item $x,y$ are represented as ${\tt uint96}$ though the right-most 18 digits are considered
as decimals.  For 96-binary bits, one can represent numbers small than 79228162514264337593543950335.
So with 18 decimal ERC20 tokens, the liquidity could contain 79 billion coins of each token which are sufficient for most tokens.  
\item Liquidity value $\Omega$: this is the inherited value ${\tt totalSupply}$ from the ERC20 and is 
of type  ${\tt uint256}$.
\item ${\tt rSquare}$ is of type ${\tt uint16}$ (where $r^2=(10000+{\tt rSquare})\cdot 10^{14}$). 
This allows us the let ${\tt rSquare}$ takes values from $1$ to  $9999$. That is, we can select the price fluctuation 
from $[0.01, 0.9999499985]$. In this paper, we use ${\tt rSquare}=6000$ as an example. That is,  $r^2=16000\cdot 10^{14}$
and the weighted price of a token (relative the weighted price of the other token) is allowed to change from $0.7745966692$
to $1.290994449$. Note that we assume the weighted price of one token relative the weighted price of the other token is 
one at the market set up time.
\item $\lambda_0,\lambda_1$ are of type ${\tt uint16}$. This implies that both $\lambda$ are smaller than
65535. In order to support tokens with smaller decimals (less than 18), we also need  to store 
$D_0=10^{18-d_0}$ and $D_1=10^{18-d_1}$ where $d_0,d_0\le 18$ are decimals of the two tokens respectively.
We use $56$-bits to store $D_0$ and 56-bits to store $D_1$.
\item $\mu$ is of type ${\tt uint56}$ where the right-most 7 digits are fractional part. 
That is, $\mu\le 7205759403.7927936$
\end{itemize}
As a summary, a 224-bits variable is used to store these parameters related to the circle as follows
\begin{center}
\begin{tabular}{ |c|c|c|c|c|c|c|}\hline
ICO(8) & $D_0$ (56)  & $D_1$ (56) & rSquare (16) & $\lambda_0$ (16)& $\lambda_1$ (16)& $\mu$ (56) \\ \hline
\end{tabular}
\end{center}
In the CoinSwap library,  the function {\tt getReservesAndmu} returns the value  ${D_0}\lambda_0\mu \| {D_1}\lambda_1\mu$
in a 256-bit variable. That is, each ${D_i}\lambda_i\mu$ takes 128-bits. It is noted that $\lambda$ is 16-bits, $\mu$ is 56-bits.
Thus, to avoid overflow, it is required that ${D_i}< 2^{56}=72057594037927936$. In other words, we have $18-d_i\le 16$
and the underlying tokens must have at least two decimals.

{\bf Remarks}: It is required that $\mu\lambda x<10^9$ and $\mu$ has at most 7 fractional digits (see the following 
data type for $\mu$). Thus we need to have $\lambda x<10^{16}$. Furthermore, the maximal value for $\lambda$ is $2^{16}-1$.
If $\mu$ takes the minimal value and $\lambda$ takes the maximal value, the maximal value that $x$ can take 
is $5^{16}=152587890625$ which is stall smaller than the supposed maximal value $79228162514$. In other words,
No overflow should happen if we require the balances for each token is smaller than the maximal value of $96$-bits.

After the market is set up, the values of $\lambda_0,\lambda_1$ are fixed for the duration of the market. 
During the computation, we need to compute $\mu\lambda_0 x$, $(\mu\lambda_0 x-10^9)^2$, $\mu\lambda_1 y$,
and $(\mu\lambda_1 y-10^9)^2$. Since we require $\mu\lambda_0 x< 10^9$ and 
$\mu\lambda_1 y<10^9$. This means that each of $\mu\lambda_0 x$ and $\mu\lambda_1 y$
can be represented by a $25+9=34$ digits (without decimal point).
That is, $\mu\lambda_0 x$  and $\mu\lambda_1 y$ can be  represented with 113-bit binary strings
(note that 25 decimal parts can be represented with a 83-bits binary string).
Thus the values  $(\mu\lambda_0 x-10^9)^2$ and $(\mu\lambda_1 y-10^9)^2$ could be held in a 256-bit binary string.

Since it is challenging to support float computation in Solidity, 
for easy calculation of (\ref{1stmu}), we can multiply $10^{36}$ to both side of the equation so that $x,y$
can then be treated as ${\tt uint}$.  That is,
$$ (\mu\lambda_0 x_0\cdot 10^{18}-10^{27})^2 + \left({\mu\lambda_1y_0\cdot 10^{18}}-10^{27}\right)^2=16000\cdot 10^{50}$$

\vskip 5pt
\noindent
{\bf Adding liquidity}.
Assuming that the current market status is ${\bf q}_0=(x_0,y_0)$ and 
the total supply of the token pair market is $\Omega_0$. If a liquidity provider would like 
to add $\Delta_x$ token $A$ coins to the market, then she/he also needs to add 
$\Delta_y= \frac{\Delta_xy_0}{x_0}$ token $B$ coins to the market. 
The new total supply of the token pair market becomes 
$$\Omega_1=\frac{\Omega_0 (\lambda_0x_1 +\lambda_1y_1)}{\lambda_0x_0 +\lambda_1 y_0}$$
where $x_1=x_0+\Delta_x$ and $y_1=y_0+\Delta_0$.
As a return, the liquidity provider 
will receive  $\Omega_1-\Omega_0$ liquidity coins for this token pair market.
The variable $\mu$ is adjusted in such a way that the following equation holds
\begin{equation}
\label{2ndmu}
(\mu\lambda_0 x_1-10^{9})^2 + \left({\mu\lambda_1 y_1}-10^{9}\right)^2=16000\cdot 10^{14}
\end{equation}
with $\mu\lambda_0 x_1<10^9$ and ${\mu\lambda_1 y_1}<10^9$.

The minted liquidity and the adjusted $\mu$ could be computed alternatively as follows. Assume 
that $x_0\not=0$. Then we have 
$\mu_0\lambda_0x_0=\mu\lambda_0 x_1$. That is, $\mu=\frac{\mu_0x_0}{x_1}$ and
$\Omega_1=\frac{\Omega_0x_1}{x_0}$.

\vskip 5pt
\noindent
{\bf Removing liquidity}.
Assuming that the current market status is ${\bf q}_0=(x_0,y_0)$ and 
the total supply of the token pair market is $\Omega_0$. If a liquidity provider would like 
to convert $\Delta_\Omega$ liquidity coins back to native coins, the liquidity 
provider will receive $(\Delta_x, \Delta_y)$ tokens where 
$\Delta_x=\frac{\Delta_\Omega x_0}{\Omega_0}$ and 
$\Delta_y=\frac{y_0\Delta_\Omega}{\Omega_0}$.
The new total supply of the token pair market becomes 
$\Omega=\Omega_0-\Delta_\Omega$.  The multiplicative scaling variable $\mu$ 
is adjusted in such a way the equation (\ref{2ndmu}) holds for 
$x_1=x_0-\Delta_x$ and $y_1=y_0-\Delta_y$ 
with $\mu\lambda_0 x_1<10^9$ and ${\mu\lambda_1 y_1}<10^9$.

\vskip 5pt
\noindent
{\bf Swapping}.
Assuming that the current market status is ${\bf q}_0=(x_0,y_0)$ and the multiplicative scaling variable is $\mu$.
For each transactions, the customer should pay 0.03\% transaction fee.  Thus if a customer
submits $\Delta_x$ token $A$ coins to the market, the customer receives $\Delta_y$ token $B$ coins such that 
\begin{equation}
\label{swapequrev1}
\left(\mu\lambda_0(x_0+0.997\Delta_x)-10^9\right)^2+\left({\mu\lambda_1 (y_0-\Delta_y)}-10^9\right)^2\le
\left(\mu\lambda_0 x_0-10^9\right)^2+\left({\mu \lambda_1y_0}-10^9\right)^2
\end{equation}
where $\mu\lambda_0(x_0+0.997\Delta_x)<10^9$.  Note that equation (\ref{swapequrev1}) is equivalent to the following equation.
\begin{equation}
\label{swapequrev11}
\left(10^{25}\mu\lambda_0(10^3\cdot x_0+997\Delta_x)-10^{37}\right)^2+\left({10^{28}\mu\lambda_1 (y_0-\Delta_y)}-10^{37}\right)^2\le
\left(10^{28}\mu\lambda_0 x_0-10^{37}\right)^2+\left({10^{28}\mu \lambda_1y_0}-10^{37}\right)^2
\end{equation}
where we try to convert the fractional parts of $\mu$ and $x_0,y_0$ into integer parts.
That is, $\mu$ has 7 fractional digits and $x_0,y_0$ have 18 fractional digits.
In order to compute the value $\Delta_y$ using the equation (\ref{swapequrev11}), let 
$$r_0=\left(10^{28}\mu\lambda_0 x_0-10^{37}\right)^2+\left({10^{25}\mu\lambda_1 y_0}-10^{37}\right)^2-
\left(10^{25}\mu\lambda_0 (1000\cdot x_0+997\Delta_x)-10^{37}\right)^2.$$
Then we need to have 
$$10^{37}-{10^{28}\mu\lambda_1 (y_0-\Delta_y)}\le \sqrt{r_0}.$$
That is,
$$10^{18}\Delta_y \le  \frac{10^{28}\mu\lambda_1y_0+\sqrt{r_0}-10^{37}}{10^3 \cdot (10^{7}\mu)\lambda_1}.$$
By the discussion in the preceding paragraphs, $10^{25}\mu\lambda_0 x_0$ and $10^{25}\mu\lambda_1 y_0$
could be represented using 113-bit binary strings. Thus  $10^{28}\mu\lambda_0 x_0$ and $10^{28}\mu\lambda_1 y_0$
could be represented using 123-bit binary strings. On the other hand, $10^{37}$
could also be represented by a 123-bit binary string. In a summary, $r_0$ could be represented
by a 247-bit binary string. Thus 
in order to get $10^{18}$ fractional digit accuracy for $\Delta_y$, it is sufficient for $\sqrt{r_0}$ to have 
accuracy until the decimal point. 

Given $\Delta_y$, in order to compute $\Delta_x$, we have the following formula:
$$r_1=\left(10^{28}\mu\lambda_0 x_0-10^{37}\right)^2+\left({10^{28}\mu\lambda_1 y_0}-10^{37}\right)^2-
\left(10^{28}\mu\lambda_1 (y_0-\Delta_y)-10^{37}\right)^2.$$
Then we need to have 
$$10^{37}-{10^{25}\mu\lambda_0 (10^3x_0+997\Delta_x)}\le \sqrt{r_1}.$$
That is,
$$10^{18}\cdot \Delta_x \ge  \frac{10^{37}-\sqrt{r_1}- 10^{28}\mu\lambda_0x_0}{997\cdot (10^7\mu)\lambda_0}.$$
Similar analysis as in the preceding paragraphs shows that,  $r_1$ has at most 247-bits and,
in order to get $10^{18}$ fractional digit accuracy for $\Delta_x$, it is sufficient for $\sqrt{r_1}$ to have 
accuracy until the decimal point. 

\subsection{Protocol fees}
\label{protfeesec}
For each transaction, traders pay 0.30\% fees on all traders. The system
collects 0.05\% protocol fee (included in the 0.30\% transaction fee) when protocol fee is turned on.
If this protocol fee is collected each time when a transaction is done, it would cost too much
gas fee. Thus we adopt the mechanism that was taken by Uniswap that this fee is only calculated
when liquidity is deposited or withdrawn or if a command for calculating protocol fee is received. 
Assume that the current market state is $(x_0, y_0)$,
the total liquidity supply amount is $\Omega$, and the current multiplicative scaling variable is $\mu_0$.
Let $\mu$ be the number such that equation (\ref{1stmu}) holds
with $\mu\lambda_0  x_0<10^9$ and ${\mu \lambda_1 y_0}<10^9$.
By equation (\ref{1stmu}), the accumulated transaction fees are 
$\frac{\mu_0 x_0-\mu x_0}{\mu_0}$ coins of token $A$ and  $\frac{\mu_0 y_0-\mu y_0}{\mu_0}$ coins of token $B$.
Thus we need to calculate the accumulated protocol fee $\Delta_\Omega$ such that 
$$\frac{\Delta_\Omega}{\Omega+\Delta_\Omega}=\frac{1}{6}\frac{\frac{(\mu_0 -\mu)\lambda_0 x_0}{\mu_0}+ 
\frac{(\mu_0 -\mu)\lambda_1 y_0}{\mu_0}}{\lambda_0 x_0+\lambda_1y_0}
=\frac{\mu_0-\mu}{6\mu_0}.$$
That is, 
\begin{equation}
\label{protocolfeeequ}
\Delta_\Omega=\frac{\Omega (\mu_0 -\mu)}{5\mu_0 +\mu}.
\end{equation}
Transfer the liquidity amount $\Delta_\Omega$ to the given protocol fee address and 
increase the total supply liquidity to $\Omega+\Delta_\Omega$. The new multiplicative scaling variable becomes $\mu$. 

\subsection{Cumulative price}
CoinSwap employs the cumulative price mechanism  for price oracle services and the smart contract records
the cumulative price ratio $P_{y/x}$ of the equation (\ref{priceratioeq}).
Let $t_0, t_1, \cdots, t_m$ be the price checkpoints where the corresponding reserves are $(x_0,y_0), \cdots, (x_m,y_m)$.
Then the smart contract records cumulative prices at time $t_m$ as the time-and-$(\lambda_0, \lambda_1)$-weighted 
average of the prices at these checkpoint times:
\begin{equation}
p_{m}=\sum_{i=1}^m\frac{\Delta_i\lambda_0(10^9-\mu\lambda_0x_i)}{\lambda_1(10^9-\mu\lambda_1y_i)}
\end{equation}
where $\Delta_i=t_i-t_{i-1}$. The $(\lambda_0, \lambda_1)$-weighted price for the time period 
$[t_{m_1}, t_{m_2}]$ is then computed as 
\begin{equation}
p_{m_1,m_2}=\sum_{i=m_1+1}^{m_2}\frac{\Delta_i\lambda_0(10^9-\mu\lambda_0x_i)}{\Delta_{m_1,m_2}\lambda_1(10^9-\mu\lambda_1y_i)}
\end{equation}
where $\Delta_{m_1,m_2} =t_{m_2}-t_{m_1}$.

\subsection{Calculation of the multiplicative variable}
\label{calmusec}
In the algorithms of  Sections \ref{addremoveliquidity} and \ref{protfeesec},
given $(x_0,y_0)$, one needs to find the multiplicative scaling variable $\mu$ such that
the equation (\ref{1stmu}) holds. 
To reduce the computation on the blockchain
and to reduce the gas fee cost, the calculation of $\mu$ could be done by the client 
(e.g., traders, liquidity providers, etc.). That is, the automated market smart contract 
only needs to verify the correctness of the value $\mu$. For the circle
\begin{equation}
\label{mueq} 
(\mu\lambda_0 x_0\cdot 10^{25}-10^{34})^2 + \left({\mu\lambda_1y_0\cdot 10^{25}}-10^{34}\right)^2=r\cdot 10^{64}
\end{equation}
and a point $(x_0,y_0)$ with $x_0+y_0\not=0$, the value of $\mu$ is computed as follows.  First the equation (\ref{mueq}) 
can be converted to the following equation
$$10^{36}\left(x_0^2\lambda_0^2+{y_0^2}{\lambda_1^2}\right)(10^7\mu)^2
-2\cdot 10^{52}\cdot \left(x_0\lambda_0+{y_0}{\lambda_1}\right)(10^7\mu)
+2\cdot 10^{68}-r\cdot 10^{64}=0.$$
That is 
\begin{equation}
\label{mucalequ}
10^7\cdot \mu=\left\lceil 10^{32}\cdot\frac{10^{20}\left(x_0\lambda_0+y_0\lambda_1\right)\pm 
\sqrt{10^{40}\left(x_0\lambda_0+y_0\lambda_1\right)^2
- 10^{36}(20000-r)\cdot\left(x_0^2\lambda_0^2+y_0^2\lambda_1^2\right)} }{10^{36}(x_0^2\lambda_0^2+y_0^2\lambda_1^2)}\right\rceil.
\end{equation}
For the two solutions in equation (\ref{mucalequ}), one uses
the $\mu$ such that $\mu\lambda_0 x_0<10^9$ and ${\mu\lambda_1 y_0}<10^9$.
That is, $\lambda_0 x_0<10^{16}$ and ${\lambda_1 y_0}<10^{16}$.
That is, the value inside $\sqrt{\cdot}$ has no fractional digits and is smaller than 
$10^{5+2(16+18)}=10^{73}$ which requires at most 245-bits.
Since the denominator in (\ref{mucalequ}) has at least 36 digits
and there is a constant scale $10^{32}$ in the numerator, 
In order to keep all digits in $10^7\cdot \mu$ significant, we need 
the square root to be approximated at least at $O(10^3)$. 
This is easily achieved using Newton's approach.

\subsection{Some practical considerations}
{\bf Calculation of $\lambda_0$ and $\lambda_1$}: 
The values of $\lambda_0$ and $\lambda_1$ are determined by the liquidity that the user
decides to put in. For example, if the user decides to put in $\frac{x}{10^{18}}$ shares of token $A$ and $\frac{y}{10^{18}}$ shares of token
$B$ where the market values of these $A$ tokens is equivalent to $B$ tokens. 
The user interface should help the user to calculate  $\lambda_0$ and $\lambda_1$ so that 
$\lambda_0 x\sim \lambda_1 y$ where $\lambda_0,\lambda_1$ are 16 bits.
Without loss of generality, we may assume that $x\ge y$. 
Let $x=x_0x_1\cdots x_m.x_{m+1}\cdots $
and $y=y_0y_1\cdots y_m.y_{m+1}\cdots$ are binary representations of $x$ and $y$. It is noted 
that $y_0$ may be zero in some cases.
The system should take $\lambda_x=y_0\cdots y_{16}$ and $\lambda_y=x_0\cdots x_{16}.$

\noindent
{\bf Circle parameters}: The default parameter for the circle is ${\tt rSqaure}=16000\times 10^{14}$.
In the router smart contracts, a user needs to provide the 
$${\tt circle}={\tt rSquare}\cdot 2^{32} + \lambda_0\cdot 2^{16}+\lambda_1$$ 
parameter for adding liquidity. For a pair of tokens $A$ and $B$, the token with smaller address is defined
as the token0. However, these information should be transparent to the end users. Thus 
the system should compare the address of token $A$ and token $B$. If token $A$ address is less than token $B$,
the system should set $\lambda_0=\lambda_x$. Otherwise, it should set $\lambda_0=\lambda_y$. 
After the value of ${\tt circle}$ is calculated, the system can automatically use this value for adding liquidity.

\noindent
It should be noted that the first transaction on each block needs to invoke the cumulative price update 
process. Thus it will pay a slightly high gas fee. The ${\tt revisemu}$ process is only invoked 
in the functions ${\tt mint}$, ${\tt burn}$, and ${\tt skim}$. After each swap transaction, the circle radius 
may decrease. Thus unless these functions are called (e.g., a  ${\tt skim}$ function call without adding/removing liquidity),
the swap transactions are carried out on a circle with smaller radius which means the price fluctuation is slightly larger--but
the difference may not be quite observable).

\noindent
{\em It is noted that a user may carry out a sequence of actions in one transaction:
make a swap to reduce the number of one token; add the liquidity based on the new coin ratios;
make another swap to cancel the first swap impact. Can a use make a profit from this? }

\section{Implementing constant ellipse based AMM -- Approach II}
\label{implapp3}
Section \ref{implappI} discusses how to implement  constant ellipse based AMM
using the circle $(x-c)^2 + (y-c)^2=r^2$ with a multiplicative scaling variable $\mu$. This section 
provides an implementation without the scaling variable $\mu$. The cost function for the AMM is defined
as $C({\bf q})=(x-10^9)^2+(y-10^9)^2$ with $x,y<10^9$. 
Each token pair market maintains a constant $\lambda_0$ which is determined 
at the birth of the market and  maintains a total liquidity supply amount variable
$\Omega$.  For the function $\pi(x,y)$ that is 
used to infer market liquidity values, we use $\pi(x,y)=\frac{x+y\lambda_0}{2}$
as in Section \ref{implappI}. As we have mentioned in Section \ref{implappI}, one may also
use $\pi(x,y)=\sqrt{x^2+y^2}$ or $\pi(x,y)=\sqrt{xy}$. But then the calculation could be more complicated.

\vskip 5pt
\noindent
{\bf Establishing a token pair market}.
When a token pair market is not established yet, a liquidity provider can establish 
the token pair market by depositing $x_0$ coins of token $A$ and $y_0$ coins of token $B$.
The assumption is that the market value of $x_0$ coins  of  token $A$ is equivalent to
the market value of $y_0$ coins of token $B$. The token pair constant variable $\lambda_0$ is defined
as $\lambda_0x_0\sim \lambda_1 y_0$. That is, at the birth of the token pair market, one coin of token
$A$ is worthy of $\lambda_0$ coins of token $B$. Let the total supply of the token pair market be 
$\Omega_0=10^{18}\cdot\frac{\lambda_0x_0+\lambda_1y_0}{2}$. 
As a return, the liquidity provider receives $\Omega_0$ liquidity coins for this token pair market. 

\vskip 5pt
\noindent
{\bf Adding liquidity}.
Assuming that the current market status is ${\bf q}_0=(x_0,y_0)$ and 
the total supply of the token pair market is $\Omega_0$. If a liquidity provider would like 
to add $\Delta_x$ token $A$ coins to the market, then she/he also needs to add 
$\Delta_y= \frac{\Delta_xy_0}{x_0}$ token $B$ coins to the market. 
The new total supply of the token pair market becomes 
$$\Omega_1=\frac{\Omega_0 (\lambda_0 x_1 +\lambda_1 y_1)}{\lambda_0x_0 +\lambda_1y_0}$$
where $x_1=x_0+\Delta_x$ and $y_1=y_0+\Delta_0$.
As a return, the liquidity provider 
will receive  $\Omega_1-\Omega_0$ liquidity coins for this token pair market.

\vskip 5pt
\noindent
{\bf Removing liquidity}.
Assuming that the current market status is ${\bf q}_0=(x_0,y_0)$ and 
the total supply of the token pair market is $\Omega_0$. If a liquidity provider would like 
to convert $\Delta_\Omega$ liquidity coins back to native coins, the liquidity 
provider will receive $(\Delta_x, \Delta_y)$ tokens where 
$\Delta_x=\frac{\Delta_\Omega x_0}{\Omega_0}$
and  $\Delta_y=\frac{\Delta_\Omega y_0}{\Omega_0}$.
The new total supply of the token pair market becomes 
$\Omega=\Omega_0-\Delta_\Omega$.  

\vskip 5pt
\noindent
{\bf Swapping}.
Assuming that the current market status is ${\bf q}_0=(x_0,y_0)$.
For each transaction, the customer should pay 0.03\% transaction fee.  Thus if a customer
submits $\Delta_x$ token $A$ coins to the market, the customer receives $\Delta_y$ token $B$ coins such that 
\begin{equation}
\label{swapequrev}
\left(\lambda_0(x_0+0.997\Delta_x)-10^9\right)^2+\left(\lambda_1({y_0-\Delta_y})-10^9\right)^2\le
\left(\lambda_0x_0-10^9\right)^2+\left(\lambda_1y_0-10^9\right)^2
\end{equation}
where $\lambda_0({y_0-\Delta_y})<10^9$.

\vskip 5pt
\noindent
{\bf Protocol fee calculation}. Protocol fees should be be transferred 
each tie when a swap transaction is made. Otherwise, we need 
to store value $\left(\lambda_0x_0-10^9\right)^2+\left(\lambda_1y_0-10^9\right)^2$ of the last time 
that the protocol fee was calculated and the calculation could be quite complicated.
Assuming that the current market status is ${\bf q}_0=(x_0,y_0)$ and 
the total supply of the token pair market is $\Omega_0$. If a client made a transaction of $\Delta_x$,
then the protocol fee is $0.003\Delta_x$. Then the protocol fee is 
$$\Omega=\frac{0.003\Omega_0\lambda_0 \Delta_x }{\lambda_0x_0 +\lambda_1y_0}$$

\section{Square roots with Babylonian/Newton method}
\label{sqr50sec}
For a given $a>0$, we need to know $x=\sqrt{a}$. We start with a guess $x_1>0$ and compute the sequences
\begin{equation}
\label{babylonianeq}
x_{n+1}=\frac{1}{2}\left(x_n+\frac{a}{x_n}\right).
\end{equation}
In other words, we use the arithmetic mean to approximate the geometric mean. This approach was used 
by the ancient Babylonians circa 1000 BC and is equivalent to the Newton's method for finding the root of 
$f(x)=0$ from a guess $x_n$ by approximating $f(x)$ as its tangent line $f(x_n)+f'(x_n)(x-x_n)$. That is,
\begin{equation}
\label{newtonite}
x_{n+1}=x_n-\frac{f(x_n)}{f'(x_n)}
\end{equation}
is a better approximation than $x_n$. If we replace $f(x)$ with $x^2-a$, we get (\ref{babylonianeq}).
A simple analysis shows that no matter whether $x_1>\sqrt{a}$ or not, we have $x_2>\sqrt{a}$.
Furthermore, we have $x_{2}> x_3> \cdots >x_{n+1} >\sqrt{a}$. By using Taylor-expansion,
it can been shown that the error roughly squares (i.e., halves) after each iteration. In other words,
the number of accurate digits approximately doubles on each iteration.  However, this convergence property
is not preserved if rounding happens. In most Solidity implementation of the Babylonian method, 
one round $x_n$ to the integer digit each time. Thus the convergence may take linear time.
The best practice is to have at least one more digit than the required accuracy.
The other challenge is that Solidity only supports ${\tt uint256}$. For an integer of 256 bits,  
the operation $a/x_n$ only returns the integer part. That is, we cannot calculate fractional part of the square roots
directly. 

In the following, we propose an approach to deal with this situation. Given $x_n=x_{n}'+\frac{x_n''}{10^l}$
where $x_{n}'$ and ${x_n''}$ are integers. Let $q=a/x'_n$ and $r=a\% x'_n$. That is,
$a=qx'_n+r$. Then we should have 
\begin{equation}
\label{diveq}
x_{n+1}
=\frac{1}{2}\left(x_n+\frac{q(x'_n+\frac{x_n''}{10^l})-\frac{qx_n''}{10^l}+r}{x_{n}'+\frac{x_n''}{10^l}}\right)
=\frac{1}{2}\left(x_n+q+\frac{r-\frac{qx_n''}{10^l}}{x_{n}'+\frac{x_n''}{10^l}}\right)
\end{equation}
In other words, we first calculate $q$ using the ${\tt uint256}$ division operator in Solidity.
Then we compute $\frac{10^lr-{qx_n''}}{10^lx_{n}'+x_n''}$. 
Let $|x|_b$ to denote the number of binary bits of an integer $x$. 
In practice, we may assume that $|a|_2\le 256$ and  take $128\ge |x_n'|_b\ge \frac{|a|_2}{2}-8$.  Then we have 
\begin{itemize}
\item $|10^lr|_b\le |x_n'|_b+|x_n''|_b\le 128+ |x_n''|_b$.
\item $|q|_b+|x_n'|_b\le |a|_2$. That is, $|qx_n''|_b\le |q|_b +|x_n''|_b\le  |a|_2 -|x_n'|_b+|x_n''|_b
\le  \frac{|a|_2}{2}+8+|x_n''|_b\le 136+ |x_n''|_b$.
\end{itemize}

Now assume that we want to have 50-bit fractional precision. Then we can set $||x_n''||_b=51$. Then we 
will be able to compute $\frac{10^lr-{qx_n''}}{10^lx_{n}'+x_n''}$ with at least 50 bits precision by
computing $2^{51}\cdot (10^lr-{qx_n''}) / (10^lx_{n}'+x_n'')$ using Solidity division.

\subsection{Montgomery algorithm for 512-bit division}
Section \ref{sqr50sec} presents an algorithm to compute square roots of a ${\tt uint256}$ with 
50-bits fractional accuracy with Solidity division. For the calculation of $\mu$ in (\ref{mucalequ}),
we need to compute the square root of a 250-bits integer with at least 33 accurate fractional digits 
(that is, 100 binary fractional bits). This can be achieved by computing the square root
of $a\cdot 10^{66}$ without accurate digit until decimal point. In other words,
we need to calculate $\frac{a*10^{66}}{x_n}$ for the iteration (\ref{babylonianeq}) where
$a$ is 250-bits and $a*10^{66}$ can only be represented as a 512-bit integer. We employed
the division algorithm described in \cite{mayer2013efficient}. The reader is referred 
to \cite{mayer2013efficient} for algorithm details. The core part of the algorithm is based on 
Newton's method (\ref{newtonite}) with $f(x)=\frac{1}{x}-a$. The iteration will be able to 
compute the reciprocal $\frac{1}{a}$ for $a$. Since $f'(x)=-\frac{1}{x^2}$, (\ref{newtonite}) could be re-written as
$$x_{n+1}=x_n-\frac{\frac{1}{x_n}-a}{-\frac{1}{x_n^2}}=x_n+(x_n-dx_n^2)=x_n(2-ax_n).$$
It is straightforward to show that the iteration has the quadratic convergence also (the error rate
squares after each iteration).

\section{Smart contracts for CoinSwap based on constant circle model}
We have implemented a proof of concept constant circle automated market (called CoinSwap)  using the approach I
in Section \ref{implappI}. The implementation is based on the Uniswap V2 architecture.
The ${\tt CoinSwapFactory.sol}$ is similar to the Uniswap Factory contracts. The major 
new component is the ${\tt CoinSwapPair.sol}$ contract. 
${\tt CoinSwapPair.sol}$ contains the following storage variables
\begin{itemize}
\item ${\tt address\ factory}$: The address who calls constructor ${\tt factory = msg.sender}$.
\item ${\tt address\  token0}$ and  ${\tt address\ token1}$.
\item ${\tt uint\ public\ priceCumulative}$ is for the cumulative price storage.
\item ${\tt uint\ public\ circleData}$ represents the following value
$${\tt ICO}\cdot 2^{216}+D_0\cdot 2^{160}+D_1\cdot 2^{104}+r\cdot 2^{88}+\lambda_0\cdot 2^{72}+ \lambda_1\cdot 2^{56}+\mu$$
where $D_0$ and $D_1$ are 56-bits each, ${\tt ICO}$ is 8-bits, $r$ is 16-bits, $\lambda_0,\lambda_1$
are 16-bits each and $\mu$ is 56-bits. The value ${\tt ICO}=0$ for regular pairs and ${\tt ICO}>0$ for initial coin offers (that is, the purpose of the 
pair is to sell ICO tokens instead of regular AMM trading).
These parameters define the constant circle 
$$(x-10^9)^2+(y-10^9)^2=(10000+r)\cdot 10^{14}.$$
 Furthermore,  
$\lambda_0 x_0\sim \lambda_1 y_0$ and $\mu$ is the  multiplicative scaling variable.
$\mu\lambda_0$ and 
$\mu\lambda_1$ are stored there to save gas cost for the multiplication during
swapping process.
\item ${\tt uint224 \ reserve}$ represents the following value
$${\tt reserve0}*2^{128}+{\tt reserve1}*2^{32}+{\tt blockTimestampLast}$$
where ${\tt reserve0}$ and ${\tt reserve1}$ are 96-bits each and ${\tt blockTimestampLast}$ is 32-bits.
\end{itemize}
${\tt CoinSwapPair.sol}$ contains the following major functions: 
\begin{itemize}
\item  ${\tt constructor()}$ sets the ${\tt factory}$ as the message sender.
\item ${\tt initialize(token0, token1, circle)}$: this function can only be called by ${\tt factory}$ and it 
initializes the two token addresses and sets the circle parameter within the ${\tt circleData}$ as
$${\tt circle}={\tt ICO}\cdot 2^{216}+D_0\cdot 2^{160}+D_1\cdot 2^{104}+r\cdot 2^{88}+\lambda_0\cdot 2^{72}+ \lambda_1\cdot 2^{56}.$$
\item ${\tt ICOmanage(unlocked, \ circleData)}$. The pair owner may be able to reset of ICO token 
selling prices. In order to call this function, it is required to have ${\tt ICO}>0$. That is, this function cannot be called 
for regular token pairs with ${\tt ICO}=0$.
\item ${\tt revisemu(balance0, balance1, } \mu)$ function re-computes the new value $\mu$ and updates it.
\item ${\tt getReserves()}$ returns ${\tt reserve0, reserve1, blockTimestampLast}$.
\item ${\tt \_update(balance0, balance1, reserve0, reserve1)}$: update cumulative price, 
let ${\tt reserve0=balance0}$, ${\tt reserve0=balance0}$.
\item ${\tt \_mintFee(reserve0, reserve1)}$: calculate and transfer protocol fee to ${\tt feeTo}$.
\item ${\tt   mint(to)}$:  
If this is for establishing for a token pair market (that is, ${\tt totalSupply}=0$),
then it calls ${\tt revisemu()}$ to calculate the value of $\mu$, $\mu\lambda_0$,$\mu\lambda_1$,
transfers the liquidity $\Omega_0$ to the liquidity provider and calls 
the ${\tt \_update()}$.  If this is for adding liquidity, it first calls ${\tt \_mintFee}$ to
transfer the accumulated protocol fee to the address ${\tt feeTo}$. 
Then it ${\tt revisemu()}$ and transfers the liquidity 
$\Omega_1-\Omega_0$ to the liquidity provider and calls the ${\tt \_update()}$.
\item ${\tt burn(to)}$ first calls ${\tt \_mintFee}$ to transfer the accumulated 
protocol fee to ${\tt feeTo}$. Then it removes the liquidity, ${\tt revisemu()}$, and 
calls ${\tt \_update()}$. 
\item  ${\tt swap(amount0Out, amount1Out, to, calldata\ data)}$:  checks 
whether equation (\ref{swapequrev11}) is satisfied. If the condition is satisfied, do the swap transaction and
call ${\tt \_update()}$
\end{itemize}

\section{Gas cost and comparison}
We compare the gas cost against Uniswap. During the implementation of CoinSwap, we find out that some of 
the optimization techniques that we used in Coinswap may be used to reduce the gas cost in Uniswap. Thus 
we compare the gas cost for Uniswap V2, our optimized version of Uniswap V2, and Coinswap.
Table \ref{tsfd}. In a summary, CoinSwap has an average 15\% gas-saving  over Uniswap V2.

\begin{table}
\caption{Gas cost Uniswap V2/Coinswap with liquidity size (40000000,10000000)}
\label{tsfd}
\begin{center}
\begin{tabular}{ |l|c|c|c| c|c|c|c|c|}\hline
function&mine()&swap()&swap()[1st] &add$\Omega$&remove$\Omega$&add ETH&full removal &partial removal\\ \hline
{\bf Uniswap V2}&141106&89894&101910&216512&140613&223074&123339&180355\\ \hline
{\bf Uniswap V2O}&132410&88224&100051&207368&97319&213930&122061&137061\\ \hline
{\bf CoinSwap}&109722&89348&96294&185442&67127&192027&98805&144283\\ \hline
{\bf Gas Saving}&22.24\%&0.61\%&5.51\%&14.35\%&31.92\%&13.92\%&19.89\%&20.00\%\\ \hline
\end{tabular}
\end{center}
\end{table}

\bibliographystyle{plain}

\end{document}